\PassOptionsToPackage{numbers,sort&compress}{natbib}
\documentclass[final,5p,times,twocolumn]{elsarticle}
\journal{Physics Letters B}

\usepackage[linkcolor=blue,
            citecolor=blue,
            urlcolor=blue,
            colorlinks=true,
            breaklinks
            ]{hyperref}

\usepackage{booktabs}
\usepackage[italic]{hepnames}
\usepackage{amsfonts}
\usepackage{amsthm}
\usepackage{bm}
\usepackage{slashed}
\usepackage{dsfont}
\usepackage{tikz}
\usepackage[dvipsnames]{xcolor}
\usepackage[shortlabels]{enumitem}
\usepackage{xspace}
\usepackage{siunitx}
\usepackage{bookmark}
\usepackage{bbm}
\usepackage{ulem}

\def\1eq#1{Eq.\nobreak\thinspace(\ref{#1})}

\def\2eqs#1#2{Eqs.\nobreak\thinspace(\ref{#1}) and\nobreak\thinspace(\ref{#2})}
\def\3eqs#1#2#3{Eqs.\nobreak\thinspace(\ref{#1}),\nobreak\thinspace(\ref{#2}) and\nobreak\thinspace(\ref{#3})}

\def\fig#1{\hyperref[#1]{Fig.\nobreak\thinspace\ref*{#1}}}
\def\figA#1{\hyperref[#1]{Fig.\nobreak\thinspace\ref*{#1}A}}
\def\figB#1{\hyperref[#1]{Fig.\nobreak\thinspace\ref*{#1}B}}
\def\figC#1{\hyperref[#1]{Fig.\nobreak\thinspace\ref*{#1}C}}
\def\figs2#1#2{\hyperref[#1]{Figs.\nobreak\thinspace\ref*{#1}}\nobreak\thinspace and\nobreak\thinspace\hyperref[#2]{\ref*{#2}}}
\def\figs3#1#2#3{\hyperref[#1]{Figs.\nobreak\thinspace\ref*{#1}},\nobreak\thinspace\hyperref[#2]{\ref*{#2}}\nobreak\thinspace and\nobreak\thinspace\hyperref[#3]{\ref*{#3}}}

\def\tab#1{\hyperref[#1]{Tab.\nobreak\thinspace\ref*{#1}}}

\def\sect#1{\hyperref[#1]{Sec.\nobreak\thinspace\ref*{#1}}}
\def\appref#1{\hyperref[#1]{App.\nobreak\thinspace\ref*{#1}}}

\def\ie{{\it i.e.}, }
\def\eg{{\it e.g.}, }

\newcommand{\be}{\begin{equation}}
\newcommand{\ee}{\end{equation}}

\newcommand{\bea}{\begin{eqnarray}}
\newcommand{\eea}{\end{eqnarray}}

\def\sz{S_{\!0}}            
\def\is{S^{-1}}             
\def\isz{S^{-1}_{\!0}}      

\def\g{\Gamma}              
\def\gt{\overline{\Gamma}}  
\def\gtz{\overline{\Gamma}_{\!0}} 

\def\s#1{{\scriptscriptstyle #1}}
\def\srm#1{{\rm{\scriptscriptstyle #1}}}

\def\MOMt{$\widetilde{\text{MOM}}$}

\DefineNamedColor{named}{MidnightBlue}  {cmyk}{0.98,0.13,0,0.43}

\begin{document}

\begin{frontmatter}
\title{\boldmath  
Real poles with opposite-sign residues in the non-perturbative quark propagator}

\author[l1]{R.~Alkofer}
\author[l2]{~~M.N.~Ferreira}
\author[l3]{~~A.S.~Miramontes}
\author[l3]{~~J.M.~Morgado}
\author[l3]{~~J.~Papavassiliou}

\affiliation[l1]{
  organization={\mbox{Institute of Physics, University of Graz, NAWI Graz, Universit\"atsplatz 5}},
  city={Graz},
  postcode={8010},
  country={Austria}
}

\affiliation[l2]{
  organization={\mbox{Instituto de Física, Universidade Federal do Rio Grande do Sul}},
  city={Porto Alegre},
  postcode={150151},
  country={Brazil}
}

\affiliation[l3]{
  organization={\mbox{Department of Theoretical Physics and IFIC, University of Valencia and CSIC}},
  city={Valencia},
  postcode={E-46100},
  country={Spain}
}

\begin{abstract}
We investigate the analytic structure of the quark propagator in the 
Landau gauge by dynamically coupling the standard gap equation to the 
non-perturbative quark-gluon vertex. Employing the full vertex basis, 
we demonstrate that for sub-GeV time-like momenta, the proper inclusion 
of the underlying dynamics leads to a pair of real poles with 
opposite-sign residues. In particular, in stark contradistinction to the 
results obtained in widely used approximations, we see no sign of complex 
conjugate poles.
This distinctive analytic structure evades conceptual shortcomings 
frequently associated with complex conjugate poles while remaining fully 
compatible with the aspects of color confinement related to positivity violation. 
Crucially, this novel behavior is governed by a dominant triplet of vertex form 
factors: the tree-level component, the anomalous chromomagnetic moment, and a 
component we label as ``spin-momentum curvature''.
By gradually tuning the individual strengths of these components, we 
demonstrate that while they contribute in distinct ways to the quark 
propagator, their joint action is vital for stabilizing the system. 
Together, they place the low-lying poles onto the real axis while 
producing a robust constituent quark mass of $350$~MeV.

\end{abstract}

\end{frontmatter}


\section{Introduction}\label{sec:intro}

During the past decades, significant progress has been achieved in mapping 
fundamental QCD correlation functions in the Euclidean domain, 
driven by a successful synergy between functional methods and 
gauge-fixed lattice simulations
~\cite{Alkofer:2000wg,Pawlowski:2003hq,Binosi:2009qm,Maas:2011se,Cloet:2013jya,Huber:2018ned,Dupuis:2020fhh,Ferreira:2023fva,Ferreira:2025anh,Huber:2025cbd}.
However, even in the most studied gauge, the Landau gauge, 
the quantitative understanding of these functions away from the 
space-like axis, crucial for both practical computations and the 
scrutiny of color confinement, remains largely incomplete; for related works, see \eg  \cite{Alkofer:2003jj,Strauss:2012dg,Frederico:2013vga, Dudal:2013yva,dePaula:2016oct, El-Bennich:2016qmb,Tripolt:2018xeo,Binosi:2019ecz,Dudal:2019gvn,Eichmann:2019dts,Li:2019hyv,Li:2021wol,Fischer:2020xnb,Horak:2021pfr,Horak:2021syv,Horak:2022myj,Eichmann:2021vnj,Horak:2022aza,Huber:2022nzs,Huber:2023uzd,Pawlowski:2024kxc,Ferreira:2026zpr}.

Within this context, exploring the analytic structure of the unquenched 
quark propagator is of fundamental importance.
Phenomenological applications have long benefited from standard 
truncation schemes which, while providing a fair description of 
various hadronic properties \cite{Maris:1999bh,Maris:2000sk,Cotanch:2003xv,Nicmorus:2008vb,Hilger:2015hka,Hilger:2014nma,Eichmann:2016yit,Mojica:2017tvh,Sanchis-Alepuz:2017mir,Weil:2017knt,Serna:2017nlr,Wallbott:2019dng,Santowsky:2020pwd,Miramontes:2021exi,Gao:2024gdj,Xu:2024fun,Hoffer:2024fgm,Huber:2025cbd,Eichmann:2025wgs,Hagel:2025ngi,Shi:2026wca}, endow the quark propagator with a 
pair of complex conjugate poles \cite{Fischer:2008sp,Windisch:2016iud,Sanchis-Alepuz:2017jjd,Miramontes:2019mco,Alkofer:2026vux}. 
These poles have been occasionally interpreted as suggestive of 
confinement ~\cite{Maris:1999nt,Chang:2009zb,Kondo:2020qdf}; nevertheless, it remains unclear whether this contentious feature represents 
an intrinsic property of the theory or an artifact of omitting subdominant 
vertex components \cite{Alkofer:2003jj,Pawlowski:2024kxc,Wieland:2026iml}. Evidently, a conclusive resolution of this question 
requires a full-fledged dynamical analysis, where the structure of the 
vertex is derived from its own underlying field equations.

In the present work, we take a significant step in this direction by solving the 
fully coupled system composed of the gap equation for the up quark and the 
corresponding Schwinger-Dyson equation (SDE) for the quark-gluon vertex. 
In the Landau gauge employed here, the latter consists of eight tensorial 
elements and their attendant form factors, which for simplicity are evaluated in 
the {\it soft-gluon limit}.
We demonstrate that the inclusion of this complete vertex 
structure converts the typical complex conjugate poles into a pair of 
real poles with residues of opposite-sign, as in the recent study of 
\cite{Wieland:2026iml}, which has been conducted within the 
quenched approximation in the chiral limit.

Crucially, this novel behavior is governed by a dominant triplet 
of vertex form factors: the transversely projected 
tree-level component (TLC) $\lambda_1$, the 
chirality-violating anomalous 
chromomagnetic moment (ACM) $\lambda_4$ \cite{Chang:2010hb,Binosi:2016wcx}, and 
the chirally symmetric $\lambda_7$, for which we coin 
the term ``spin-momentum curvature'' (SMC) 
due to its explicit dependence on the 
momentum commutator $[\slashed{p}, \slashed{k}]$.\footnote{Here $p$ and $k$ denote 
the two quark momenta, see below. 
The form factor correspondence with \cite{Wieland:2026iml} is $\Gamma^{(1)} \!\!\!\longleftrightarrow {\rm TLC}$, 
$\Gamma^{(6)}  \!\!\!\longleftrightarrow {\rm ACM}$, while 
the SMC is a linear combination of $\Gamma^{(1)}$, $\Gamma^{(2)}$, and $\Gamma^{(4)}$.} 
By systematically tracking 
the migration of the propagator poles, we expose a delicate balance within 
this key triplet of form factors. While the 
TLC with unchanged strength\footnote{Note that in all studies based on the 
rainbow-ladder truncation, the strength of the TLC is either phenomenologically determined or artificially enhanced.} alone yields 
real poles but underestimates the constituent mass ($\sim 180$~MeV), the 
inclusion of the ACM at full strength produces substantial mass enhancement 
($\sim 550$~MeV) but generates complex conjugate poles. At this point, the inclusion 
of the SMC 
leads to the restoration of the 
initial pole structure. Specifically, 
by introducing a scaling parameter to gradually dial in the 
strength of the SMC, we reveal a critical threshold past which the 
complex conjugate 
poles are diverted back onto the real axis. At their full physical strengths, 
the TLC, ACM, and SMC perfectly balance the system, yielding real poles while generating a robust constituent quark mass of 
$\sim 350$~MeV. 
Crucially, on the space-like side, this complete vertex structure alters 
the momentum dependence of the mass function compared with simpler vertex models. 
This modification brings the resulting constituent mass into much closer 
agreement with the recent precise lattice results of~\cite{Oliveira:2025boh}.

The final upshot of this analysis is that the complex conjugate poles seem to 
emerge because the gap equation kernel is unbalanced w.r.t.\ the 
chiral symmetry respecting TLC, the chirality violating ACM, and 
the chiral symmetry preserving SMC. 
Specifically, when the required strength is concentrated within a restricted  
vertex structure, the geometry of the propagator poles is destabilized. 
 Instead, when the kernel support is properly distributed among all relevant 
 form factors, as dictated by the full dynamics, 
 ({\it i})~the complex conjugate poles 
disappear, ({\it ii})~the correct mass scale is reproduced, and ({\it iii})~the momentum 
dependence of the constituent quark mass function is favorably modified.

\section{Quark propagator and quark-gluon vertex}\label{sec:sdes}

In this section we provide a concise overview of the principal 
components and general notation employed throughout the present work. Note that, in general, we formulate the main equations in Minkowski space, and recast them into Euclidean hyperspherical coordinates for the final numerical treatment.

\subsection{The basic equations}\label{sec:gen}

We work in the Landau gauge, where the gluon propagator, \mbox{$\Delta^{ab}_{\mu\nu}(q)=-i\delta^{ab}\Delta_{\mu\nu}(q)$}, assumes a completely transverse form
\begin{align}\label{eq:gluonprop}
\Delta_{\mu\nu}(q)&= P_{\!\mu\nu}(q)\Delta(q^2)\,,& P_{\!\mu\nu}(q)&=g_{\mu\nu}-q_\mu q_\nu/{q^2}\,.
\end{align}

\begin{figure}[!t]
\centering
\includegraphics[scale=0.95]{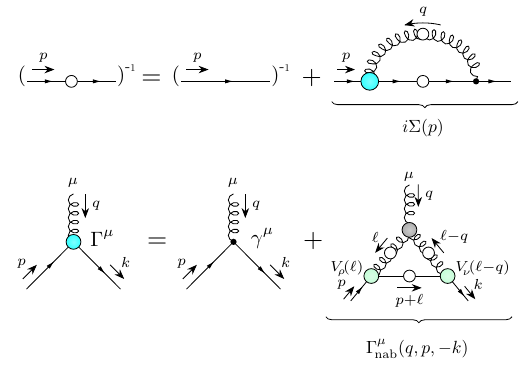}
\caption{Upper panel: Diagrammatic representation of the quark gap equation. Lower panel: The SDE of the quark-gluon vertex in the SV approximation.
White circles denote full propagators, the cyan circle represents the full quark-gluon vertex,  the green circles indicate the combination  $V(q^2)\gamma^{\mu}$, while the grey circle stands for the full three-gluon vertex.  }
\label{fig:sdesys}
\end{figure}

We denote the up quark propagator by \mbox{$S^{ab}(p)\!=\!i\delta^{ab}S(p)$}~\cite{Itzykson:1980rh}, and employ the standard parametrization 
\begin{align}\label{eq:invs}
S(p)&=\sigma_{\!\s{V}}(p^2)\slashed{p}+\sigma_{\!\s{S}}(p^2)\,,&\is(p)&=A(p^2)\slashed{p}-B(p^2)\,,
\end{align}
with
\begin{align}
\sigma_{\!\s{V}}(p^2) =
\frac{1}{A(p^2)[p^2 - {\cal M}^2(p^2)]} \,,
\enskip
\sigma_{\!\s{S}}(p^2) = {\cal M}(p^2) \sigma_{\!\s{V}}(p^2)\,, 
\label{eq:thesigmas}
\end{align}
where ${\cal M}(p^2) = B(p^2)/A(p^2)$ is  
the renormalization-group invariant constituent quark mass. 
The momentum evolution of these functions 
is governed by the gap equation, shown in the upper panel of  \fig{fig:sdesys}, which reads
\be\label{eq:gap}
\is(p)=\isz(p)-i\Sigma(p)\,,\
\ee
where $\sz(p)$ denotes the tree-level quark propagator, 
while 
\be\label{eq:selfenerg}
\Sigma(p)=-\frac{4g^2}{3}\int_q\gamma_{\!\mu}S(p+q)\,\gt^{\,\mu}(q,p,-p-q)\Delta(q^2)\,,
\ee
is the quark self-energy, $g$ is the QCD gauge coupling, and \mbox{$\int_{q} := (2\pi)^{-4} \int \!\!{\rm d}^4 q$} represents a suitably regularized integration over virtual momenta. For later convenience we define \mbox{$p^2\Sigma_\srm{A}(p^2)=\textrm{Tr}[\slashed{p}\Sigma(p)]$} and \mbox{$\Sigma_\srm{B}(p^2)=\textrm{Tr}[\Sigma(p)]$}. 

The key element entering in the self-energy $\Sigma(p)$ of \1eq{eq:selfenerg}
is the transversely projected quark-gluon vertex, defined as 
\mbox{$\gt^{\,\mu}(q,p,-k)=P^{\mu\nu}(q)\g_{\!\nu}(q,p,-k)$}, whose  
tree-level expression  
is given by  
$\gtz^{\,\mu}(q)=P^{\mu\nu}(q)\gamma_\nu$. 

In general kinematics, \mbox{$\overline{\Gamma}^{\,\mu}(q,p,-k)$} is spanned by eight tensor structures \cite{Aguilar:2024ciu}, 
\begin{align}\label{eq:decomp}
    \gt^{\,\mu}(q,p,-k)&=\sum_{i=1}^{8}\lambda_i(q,p,-k)\bar{\tau}_{i}^\nu(p,-k)\,,
\end{align}
where \mbox{$\bar{\tau}_{i}^\mu(p,-k)=P^\mu_{\!\nu}(q)\tau_{i}^\nu(p,-k)$}, and the  tensor basis is given by \mbox{$\tau_1^\nu=\gamma^\nu$}, \mbox{$\tau_2^\nu=(k+p)^\nu$}, \mbox{$\tau_3^\nu=(\slashed{k}+\slashed{p})\gamma^\nu$}, \mbox{$\tau_4^\nu=(\slashed{k}-\slashed{p})\gamma^\nu$},
\mbox{$\tau_5^\nu=(\slashed{k}-\slashed{p})(k+p)^\nu$}, \mbox{$\tau_6^\nu=(\slashed{k}+\slashed{p})(k+p)^\nu$}, 
\mbox{$\tau_7^\nu=-\frac{1}{2}[\slashed{k},\slashed{p}]\gamma^\nu$}, \mbox{$\tau_8^\nu=-\frac{1}{2}[\slashed{k},\slashed{p}](k+p)^\nu$}.

\begin{figure*}[!t]
\centering
\includegraphics[width=\textwidth, keepaspectratio]{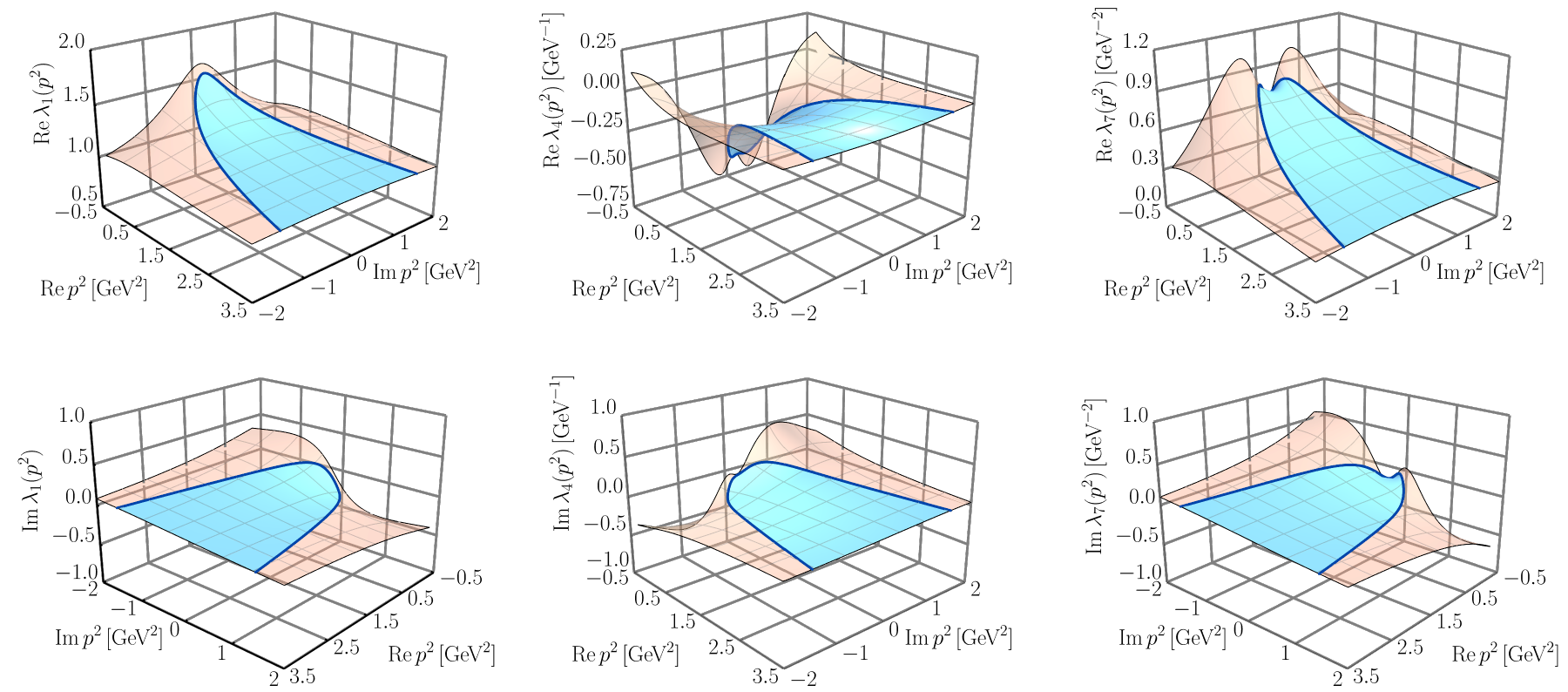}
\caption{Real (top) and imaginary (bottom) parts of the quark-gluon vertex 
form factors $\lambda_1(p^2)$, $\lambda_4(p^2)$, and 
$\lambda_7(p^2)$; displayed in the left, middle and right panels, 
respectively. The blue parabolas delimit the region of validity of the Euclidean 
formulation, where the numerical 
calculations have been performed \cite{Ferreira:2026zpr}.
The pale orange surfaces denote the result of the 
extrapolation by means of the Schlessinger Point Method (SPM).}
\label{fig:ffs}
\end{figure*}

Note that, since 
\mbox{$\bar \tau_4^\mu =P^\mu_\nu(q)(\slashed{k}-
\slashed{p})\gamma^\nu=P^\mu_\rho(q)i\sigma^{\rho\nu}q_\nu 
= i\sigma^{\mu \nu}q_\nu$}, 
the component $\lambda_4 (q,p,-k)$
corresponds to the ACM, 
a tensorial 
quark-gluon coupling akin to the loop-induced Pauli term in QED.

In addition, as we will see in \sect{sec:thetriad}, the element   
$\lambda_7 \,\bar{\tau}^\nu_{7}$ affects crucially 
the analytic structure of the quark propagator.
The commutator $[\slashed{k},\slashed{p}]$ in $\tau_7^\nu$
acts as a measure of local geometric curvature in the space of the 
quark and anti-quark momenta. By virtue of 
\mbox{$\bar \tau^\mu_7 = - \frac{1}{2}P^\mu_\nu(q)
[\slashed{k},\slashed{p}]\gamma^\nu= -i
P^\mu_\nu(q)\sigma^{\alpha\beta}k_\alpha p_\beta\gamma^\nu$}, 
this curvature structure is mapped directly onto the spin tensor 
$\sigma^{\alpha\beta}$, linking the spatial distortion to the internal 
orientation of the particle. We will therefore refer to this vertex 
component as ``spin-momentum curvature'' (SMC). 
This terminology 
is further supported by the fact that the SMC contains three Dirac matrices, 
constituting a higher-rank tensor within the Clifford algebra (corresponding 
to the basis tensor $R^{(2),\mu}$ in~\cite{Wieland:2026iml}). Consequently, the 
SMC is structurally distinct from the other three chirally symmetric form 
factors, $\bar \tau^\nu _{1,5,6}$, which contain only a single Dirac matrix 
and therefore represent standard vector-like quark-gluon couplings.

\subsection{The SDE of the quark-gluon vertex }\label{sec:vertsde}
The momentum evolution of the 
$\lambda_i(q,p,-k)$ 
is controlled by the quark-gluon vertex SDE. The most complete
 ``one-loop dressed'' form of this SDE 
\cite{Alkofer:2008tt,Williams:2015cvx,Aguilar:2024ciu,Wieland:2026iml} has been obtained within the 
three-particle irreducible (3PI) effective action formalism 
\cite{Cornwall:1973ts,Cornwall:1974vz,Berges:2004pu,Carrington:2010qq,Williams:2015cvx}, and contains  
two ``one-loop dressed'' diagrams, 
traditionally referred to as ``Abelian'' and ``non-Abelian'', see \eg Fig.\,2 in \cite{Wieland:2026iml}. 
Since it is well-known that the 
contribution of the ``Abelian'' graph is rather negligible, in what follows we will only consider the ``non-Abelian'' term.

This 3PI-SDE will be further simplified, 
motivated by the recent analysis 
of \cite{Miramontes:2025imd,Ferreira:2025wpu,Ferreira:2026gbe}. 
Specifically, 
all quark-gluon vertices inside  
the "non-Abelian" diagram are replaced by a single form factor, proportional to 
$\gamma_{\mu}$, which 
depends solely on the 
momentum of the incoming gluon. Thus,  
inside the diagram 
$\overline{\Gamma}_{\!\textrm{nab}}^{\,\mu}$
in \fig{fig:sdesys} we substitute  
\mbox{$\g^{\,\rho}(\ell,p,-\ell_1) \to V(\ell^2) 
\gamma^{\,\rho}$} and 
\mbox{$\g^\nu(-\ell_2,\ell_1,-k) \to 
V(\ell_2^2) 
\gamma^\nu$}
 where $\ell_1= p+\ell$ \, and \,  $\ell_2=\ell-q$. 

The function 
$V(q^2)$ is determined by first 
solving the 3PI-SDE 
maintaining 
full quark-gluon vertices inside 
$\overline{\Gamma}_{\!\textrm{nab}}^{\,\mu}$, and isolating 
the form factor 
$\lambda_1^{\rm \s{3PI}}(q,p,-k)$,
multiplying the classical tensor 
$\gamma_{\mu}$. Then,  
the slice of  $\lambda_1^{\rm \s{3PI}}(q,p,-k)$ corresponding to the 
symmetric configuration, \mbox{$q^2=p^2=k^2$},  is singled out,
and identified with $V(q^2)$, \, 
\ie 
\mbox{$V(q^2) = \lambda^{\rm \s{3PI}}_1 (q^2,q^2,q^2)$}.

Under this approximation,  
the vertex SDE, shown in the lower
panel of \fig{fig:sdesys},  
is given by 
\be\label{eq:vertexsde}
\overline{\Gamma}^{\,\mu}(q,p,-k)=\overline{\Gamma}^{\,\mu}_{\!0}(q)+\overline{\Gamma}_{\!\textrm{nab}}^{\,\mu}(q,p,-k)\,,
\ee
with  
\be\label{eq:c2}
\overline{\Gamma}_{\!\textrm{nab}}^{\,\mu}\!=\!\frac{3ig^2}{2}\!\!\!\int_\ell 
\overline{\Gamma}^{\,\nu}_{\!0}(\ell_2)S(\ell_1)\overline{\Gamma}^{\,\rho}_{\!0}(\ell)\overline{\Gamma}^{\,\mu}_{\nu\rho}V(\ell_2^2)\Delta(\ell_2^2)V(\ell^2)\Delta(\ell^2)\,\,,
\ee
where \mbox{$\overline{\g}^{\,\mu\nu\rho}(q,r,p)=P_{\!\!\mu'}^\mu(q)  P_{\!\!\nu'}^{\nu}(r)  P_{\!\!\rho'}^\rho(p) 
\g^{\,\mu'\! \nu' \!\rho'}(q,r,p)$}
denotes the 
transversely-projected three-gluon vertex.
According to various studies~\cite{Eichmann:2014xya,Blum:2014gna,Huber:2016tvc,Huber:2020keu,Pinto-Gomez:2022brg,Aguilar:2023qqd,Pinto-Gomez:2024mrk},  an excellent approximation to $\overline{\g}^{\, \mu \nu \rho}(q,r,p)$ may be obtained 
by setting 
\be\label{eq:planardeg}
\g^{\mu\nu\rho}(q,r,p)=L_{sg}(s^2)\left[g^{\nu\rho}(r-p)^\mu+g^{\mu\rho}(p-q)^\nu+g^{\mu\nu}(q-r)^\rho\right]\,,
\ee
with \mbox{$\displaystyle s^2=(q^2+r^2+p^2)/2$}, and the form factor $L_{sg}(s^2)$ given by Eq.~(A1) and Tab.~II of \cite{Aguilar:2023mam}.

Within this approximation, 
the final expressions for the eight 
$\lambda_i(q,p,-k)$ 
are straightforwardly obtained 
from \1eq{eq:c2}, after 
suitable projection and subsequent 
integration (no iterative procedure required).

\begin{figure*}[!t]
\centering
\includegraphics[width=1.0\textwidth, keepaspectratio]{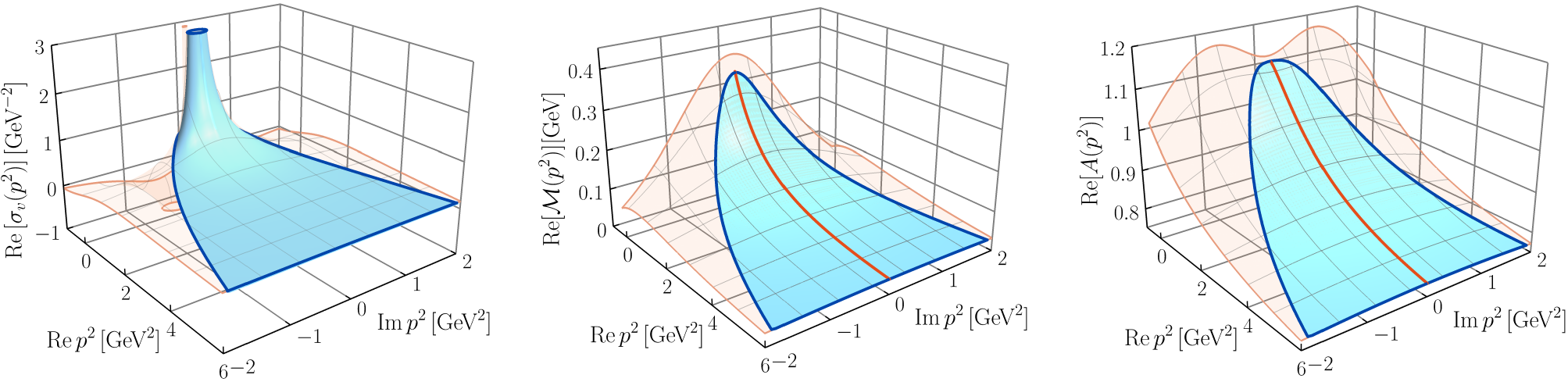}
\caption{Plots of the real parts of the Dirac component
$\sigma_{\!\s{V}}(z)$ (left panel), the constituent quark mass function, $\mathcal{M}
(p^2)$ (middle panel), and the dressing function $A(p^2)$ (right panel) in 
the complex momentum plane; all components of the quark-gluon vertex were
included in their derivation. The blue parabolas delimit the region where numerical 
calculations were performed, while the pale orange surfaces represent the SPM 
extrapolations of the corresponding results.}
\label{fig:sigma_v}
\end{figure*}

\subsection{The soft-gluon limit of the quark-gluon vertex }\label{sec:softgluon}
We simplify the analysis further, by considering the 
so-called ``soft-gluon ({\it sg})''  limit of the quark-gluon vertex \cite{Ferreira:2026zpr}, namely
\be\label{eq:vertexsg}
\gt^{\,\mu}(q,p,-k)=\sum_{i=1}^8\lambda_{i}^{sg}(p^2)\,P_{\!\nu}^\mu(q)\,\tau_i^\nu(p,-k)\,.
\ee
Note that we maintain the tensorial structure {\it intact}, while the attendant form factors $\lambda_i^{sg}(p^2):=\lim_{q\to 0}\lambda_i(q,p,-k)$ 
depend on a single kinematic variable, namely $p^2$.

The equations that determine the $\lambda_i^{sg}(p^2)$ can be derived
by taking the $q\to 0$ limit 
of the corresponding projections of the SDE in \1eq{eq:vertexsde}, namely 
\be\label{eq:sdelisg}
\lambda_i^{sg}(p^2)=\delta_{i1}+\underbrace{\lim_{q\to 0}\textrm{Tr}\left[\mathcal{P}_{\!i,\mu}(q,p,-k)\gt^{\,\mu}_{\!\textrm{nab}}(q,p,-k)\right]}_{\lambda_{i,\srm{Q}}^{sg}(p^2)}\,,
\ee
where the 
projectors $\mathcal{P}_{\!i}^{\,\mu}(q,p,-k)$ are given in 
Eq.~(3.9)~of~\cite{Aguilar:2024ciu}. 
The explicit integral expressions obtained through this procedure are given in 
Eq.~(3.9) of \cite{Ferreira:2026zpr}.

Note that, within this approximation, an ambiguity arises
when $\lambda_i^{sg}$ is 
inserted in the gap equation:
the argument of $\lambda_i^{sg}$ 
could be 
considered to be either $p^2$ or $(p+q)^2$. 
Evidently, if the momentum $p^2$ is used, the 
 $\lambda_i^{sg}(p^2)$ would appear outside the 
$q$-integration in \1eq{eq:selfenerg}. Instead, 
we choose the representative momentum to be 
$(p+q)^2$, such that the strength of $\lambda_i^{sg}((p+q)^2)$ is integrated over  
inside the gap equation, as happens when the full momentum dependence is retained.
This choice, in turn, guarantees the ultraviolet finiteness of the results, once the subtractions dictated by the renormalization procedure have been duly carried out, as described below.

\subsection{Renormalization}\label{sec:renorm}
The renormalization of  
the system of equations composed of  
\2eqs{eq:gap}{eq:vertexsde}
is carried out according to the   
procedure outlined in Sec.~V of \cite{Ferreira:2025wpu}.  
In particular, after the introduction of the 
appropriate renormalization constants (see 
\eg Eq.\,(3.13)  in \cite{Aguilar:2024ciu}),
we employ the version of the
momentum-subtraction scheme~\cite{Celmaster:1979km,Hasenfratz:1980kn,Braaten:1981dv} known as \MOMt{}~\cite{Skullerud:2002ge,vonSmekal:2009ae,Kizilersu:2021jen,Aguilar:2023mam,Aguilar:2024ciu}. This scheme is defined using as reference 
precisely the soft-gluon limit of the quark-gluon vertex, 
namely through the prescriptions \cite{Skullerud:2002ge}: 
$\Delta^{-1}_{\srm{R}}(\mu^2) = \mu^2$, 
$A_{\srm{R}}(\mu^2) = 1$, $B_{\srm{R}}(\mu^2) = m_{\srm{R}}(\mu^2)$ and $\lambda_{1, \srm{R}}^{sg}(\mu^2)=1$.
In addition, the multiplicative 
renormalization of the quark self-energy is
implemented through the expedient employed in \cite{Ferreira:2025wpu}
(see also \cite{Fischer:2003rp,Aguilar:2010cn,Aguilar:2018epe}),
leading to the effective replacement 
\mbox{$Z_1 \gamma_\mu \to V_\mu (q)$}, where 
$Z_1$ is the renormalization constant 
of $\gt_\mu(q,p,-k)$.

Thus, one arrives at the renormalized system 
of equations 
\bea
\is_{\!\srm{R}}(p) & = & \slashed{p}-m_{\srm{R}}-i\left[\Sigma_{\srm{R}}(p^2)-\Sigma_{\srm{R}}(\mu^2)\right]\,,
\nonumber\\
\lambda_{i,\srm{R}}^{sg}(p^2) & = &\lambda_{i,\srm{Q}}^{sg}(p^2)+\left[1-\lambda_{i,\srm{Q}}^{sg}(\mu^2)\right]\delta_{i1}
\label{eq:vertexren}\,,
\eea
with $\lambda^{sg}_{i,\srm{Q}}(p^2)$ defined in \1eq{eq:sdelisg}, and
\be
\Sigma_{\srm{R}}(p^2)=-\frac{4g^2}{3}\!\!\int_q \gamma_\mu\!V(q^2) S_{\!\srm{R}}(p+q)\gt_{\!\srm{R}}^{\,\mu}(q,p,-p-q)\Delta_\srm{R}(q^2)\,.
\ee

\section{Solving the coupled system in the complex plane}\label{sec:sdesys}

In this section we 
focus on the treatment of the system given by \1eq{eq:vertexren}
in the complex plane. 
We emphasize that this problem  
contains a single external momentum, namely 
$p$, which enters both in the gap equation
(first of \1eq{eq:vertexren}), as well as 
in the vertex form factors $\lambda^{sg}_i(p^2)$
(second of \1eq{eq:vertexren}). It is this particular momentum that will be complexified, $p^2 \to z$, 
while the virtual (integration) momenta 
remain real (Euclidean). Note in particular, 
that the momentum $p$ may be 
channeled exclusively 
through the 
quark propagator of the graph 
$\overline{\Gamma}_{\!\textrm{nab}}^{\,\mu}$ 
in \fig{fig:sdesys}; therefore, 
all other components of $\overline{\Gamma}_{\!\textrm{nab}}^{\,\mu}$ are evaluated at Euclidean momenta, where lattice results are available. 

\begin{figure*}[!t]
\centering
\includegraphics[width=\textwidth]{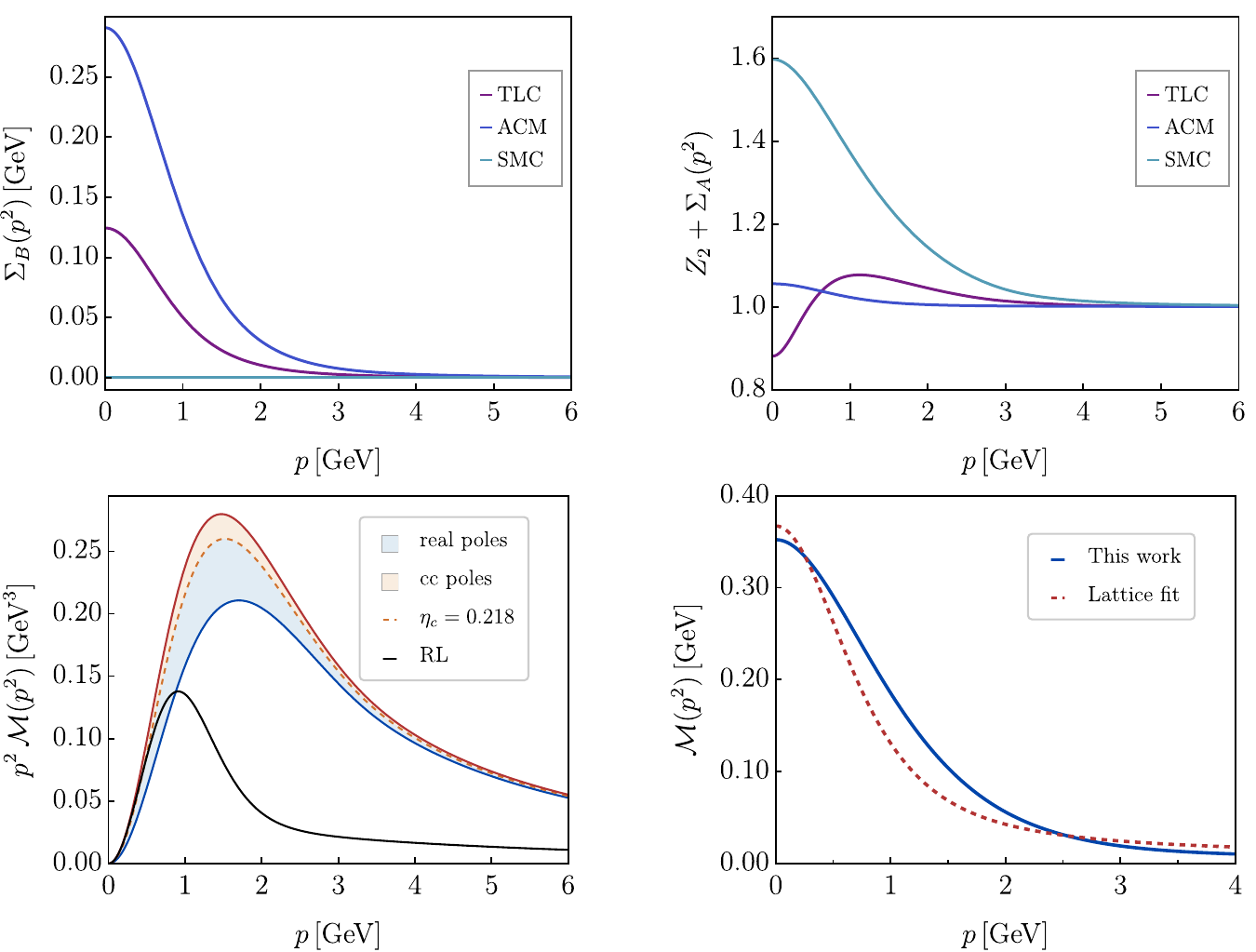}
\caption{
Upper panels: Individual contributions of the TLC, ACM, and SMC 
to the quark propagator functions $B(p^2)$ and $A(p^2)$, obtained by solving the 
gap equation with only the respective vertex component retained. The upper left 
panel clearly illustrates that the dominant contribution to $B(p^2)$---and thus to 
dynamical mass generation---originates from the ACM, while the SMC does not 
contribute to $B(p^2)$ at all. Conversely, the largest contribution to the dynamical 
dressing of $A(p^2)$ is driven by the SMC. Together, the ACM and SMC induce a 
distinct momentum dependence in the mass function ${\cal M}(p^2)$ compared to 
the TLC-only result. Lower left panel: The product $p^2 {\cal M}(p^2)$ compared 
with the corresponding rainbow-ladder (RL) truncation (black line); the solid 
red line isolates the TLC and ACM structures, while the solid blue line displays 
the full triad with the SMC included. Lower right panel: Comparison of the full 
triad result with the lattice parametrization of~\cite{Oliveira:2025boh}.}
\label{fig:dress}
\end{figure*}

\subsection{Inputs}\label{sec:inputs}

The solution of the system \1eq{eq:vertexren} requires four external ingredients, all of which are defined at $\mu=2~\textrm{GeV}$.

($i$) For the Landau-gauge gluon propagator, we employ the fit to the $N_f=2$ lattice data of \cite{Ayala:2012pb, Binosi:2016xxu}, given in Eq.~(A1) of \cite{Aguilar:2023mam} and shown in the top-left panel of Fig.~6~in~\cite{Ferreira:2025wpu}.

($ii$) For the form factor $L_{sg}(s^2)$, we use the fit to the lattice data of \cite{Aguilar:2019uob} reported in Eq.~(A1) of \cite{Aguilar:2023mam}. The corresponding form factor is displayed in the top-right panel of Fig.~6~in~\cite{Ferreira:2025wpu}.

($iii$) For the function $V(q^2)$ we adopt the parametrization given in Eq.~(6.5) of \cite{Ferreira:2025wpu}, shown in the bottom-right panel of Fig.~6 therein.

($iv$) Finally, for the renormalized current quark mass we employ $m_{\srm{R}} = 50$ MeV; we will comment on this choice in \sect{sec:results}.

\subsection{Details of the numerical treatment}\label{sec:iter}

The system of equations given by \1eq{eq:vertexren} is reformulated in Euclidean space following standard conversion rules; see \eg Sec.~IV. B and App.~A in \cite{Aguilar:2024ciu}. 
Then, the external momentum is
complexified, employing the parametrization \mbox{$p^2\to z=x+iy$}, while the integration (loop) momenta remain real and positive.

\vspace{0.2cm}

Our numerical procedure 
solves simultaneously for the $A(p^2)$, $B(p^2)$ and $\lambda^{sg}_i(p^2)$,
by treating them as a system of coupled 
equations, iterating until overall convergence is achieved. 

Specifically, the numerical algorithm is organized as follows:

({\it i}) In order to commence 
the procedure, an initial expression for the quark propagator that appears in 
the integral of \1eq{eq:c2}
is required. We find it convenient to 
employ as our initial ``guess'' function
the solution of the gap equation 
obtained in  \cite{Gao:2024gdj}; the corresponding 
$A(p^2)$ and $\mathcal{M}(p^2)$ 
are shown in Fig.~4 therein. 
Solutions obtained from other standard versions of the gap equation 
should be equally appropriate for initializing the calculation. 

({\it ii}) 
Next, we determine 
the location of the first singularities of the input propagator 
in the complex plane, denoted by \mbox{$p^2_\srm{S}=R_\srm{S} e^{i\theta_\srm{S}}$}. These singularities define the boundary of the region where the standard Euclidean formulation remains valid~\cite{Ferreira:2026zpr}, and consequently the maximal domain where the vertex form factors can be computed. This region corresponds to the parabolic domain defined through the equation 
\mbox{$y^2=a\,x+ b$}\,,
with \mbox{$a=4R_\srm{S}\cos^2(\theta_\srm{S}/2)$} and \mbox{$b=4R_\srm{S}^2\cos^4(\theta_\srm{S}/2)$}.

({\it iii}) The form factors $\lambda_i^{sg}(p^2)$ are then computed within the allowed domain through direct numerical integration.

({\it iv}) The resulting  $\lambda_i^{sg}(p^2)$ are substituted into the gap equation, which is solved for complex external momenta distributed along the boundary of the same parabolic domain while retaining the full kinematic structure of $\gt_\mu$, given by \1eq{eq:vertexsg}.  

({\it v}) Steps ({\it ii})--({\it iv}) are 
repeated until convergence has been reached simultaneously for $A(p^2)$, $B(p^2)$, and the $\lambda^{sg}_i(p^2)$, with a numerical tolerance of $10^{-6}$.

Following this algorithm, and using standard numerical techniques (see \eg \cite{Sanchis-Alepuz:2017jjd}), the quark propagator and quark-gluon vertex were determined within the domain bounded by the parabola \mbox{$y^2=a\,x+b$}, with \mbox{$a=0.6~\textrm{GeV}^2$} and \mbox{$b=0.09~\textrm{GeV}^4$}. 

\section{Results}\label{sec:results}

In this section we present a summary of the 
most prominent results of our analysis. In what follows we suppress the index ``$sg$'' throughout.

\subsection{The vertex form factors in the complex plane}\label{sec:lambdas}

The soft-gluon limit of the eight form factors $\lambda_i$ has been computed within the domain defined by the parabola mentioned above. 
In \fig{fig:ffs} we
display the 
corresponding results for \mbox{$\lambda_1(p^2)$}, \mbox{$\lambda_4(p^2)$} and \mbox{$\lambda_7(p^2)$}; this particular subset has been selected because it 
constitutes the dominant triplet of form factors that largely determines the 
bulk of the quark propagator (see \sect{sec:thetriad}). 

Note that the above results 
agree rather well with those obtained in \cite{Ferreira:2026zpr}, where a fixed propagator with complex conjugate poles was employed, instead of the dynamical quark propagator (with real poles, see \sect{sec:realpoles}) 
used here. This suggests that the form of the $\lambda_i(p^2)$ is relative insensitive 
to the details of the quark propagator
used for their computation.

\subsection{Quark propagator: Real poles with opposite-sign residues }\label{sec:realpoles}

The most noteworthy feature of the 
quark propagator that emerges from this 
analysis is the absence of complex conjugate 
poles from its Dirac components 
$\sigma_{\!\s{V}}(p^2)$ and 
$\sigma_{\!\s{S}}(p^2)$, introduced 
in \1eq{eq:thesigmas}. 
This finding is in stark contradistinction to 
what happens within the RL and related approximations, where complex conjugate poles constitute a standard feature. 
Instead, in the present case, one encounters 
a pair of real poles, denoted by 
$p^2_{1}$ and $p^2_{2}$, lying on the 
time-like axis, namely  
\begin{align}\label{eq:polepos}
p^2_{1}&=-0.16\textrm{ GeV}^2\,,& p^2_{2}&=-0.58\textrm{ GeV}^2\,.
\end{align}
Note in particular that the value of $p_1^2$ is in agreement with the 
estimate $p_1^2 \approx 1.2 {\cal{M}}^2(0)$ provided in \cite{Wieland:2026iml}. We do not find any other singularities for either 
the real or imaginary part being smaller than 1 GeV$^2$, neither in $\sigma_{\!\s{V}}(p^2)$ nor in  
$\sigma_{\!\s{S}}(p^2)$.

Quite importantly, the two poles 
have residues with {\it opposite signs}:
the first pole has a positive residue, whereas the second a negative one, namely
\begin{align}\label{eq:poleres}
\textrm{Res}(\sigma_{\!\s{V}},p^2_{1})&=1.12\,,& \textrm{Res}(\sigma_{\!\s{V}},p^2_{2})&=-0.51\,.
\end{align}

In order to establish that the 
singularities found at 
$p^2_i$ correspond to simple poles, we considered the 
\mbox{$\lim_{z\to p^2_i}(z-p^2_i)^mf(z)$},
as $m$ is varied. A careful analysis reveals that, at a high level of accuracy, 
this limit is finite for $m=1$, while 
it vanishes or diverges as $m$ moves to the 
right or to the left of 
unity, respectively.

\begin{figure}[!t]
\hspace*{-0.4cm}
\includegraphics[width=0.5\textwidth]{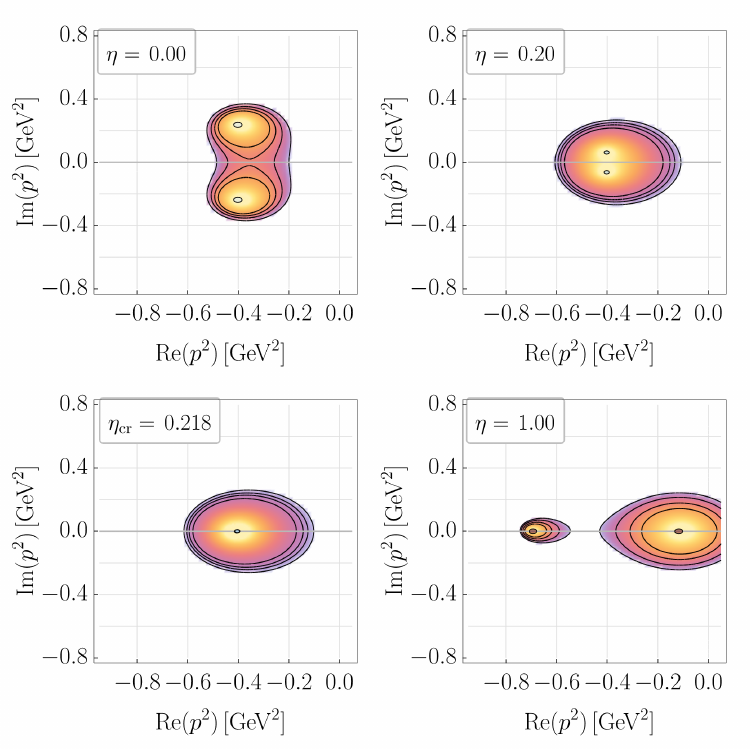}
\caption{Contour plot of $\sigma_{\!\s{V}}(p^2)$ showing the evolution of 
the pole structure as the coefficient $\eta$ is varied for the vertex in 
\1eq{eq:geff}. For small $\eta$, the propagator features a pair of complex
conjugate poles, which become real as $\eta$ exceeds a critical value $\eta_{cr}$.}
\label{fig:poles_eta}
\end{figure}

The location of the first pole, $p^2_{1}$, is determined directly from the numerical solution of the gap equation in the complex plane; it is identified with the point where the iterative procedure ceases to converge. 
The existence of this pole is independently confirmed using SPM extrapolations, constructed from the data available up until that point. This SPM analysis predicts, in addition, a second pole, namely $p^2_{2}$. 

In the left panel of \fig{fig:sigma_v} 
we show the real part of $\sigma_{\!\s{V}}(p^2)$.  The light-blue area indicates the domain defined by the parabola where direct numerical calculations were performed, up to the point where the 
first pole, $p_1^2$, was reached. 
The pale orange surface represents the SPM extrapolation beyond this region; the 
appearance of a second pole, $p_2^2$,  pointing towards the opposite direction, is marginally visible.

\begin{figure*}[!t]
\centering
\includegraphics[width=0.9\textwidth]{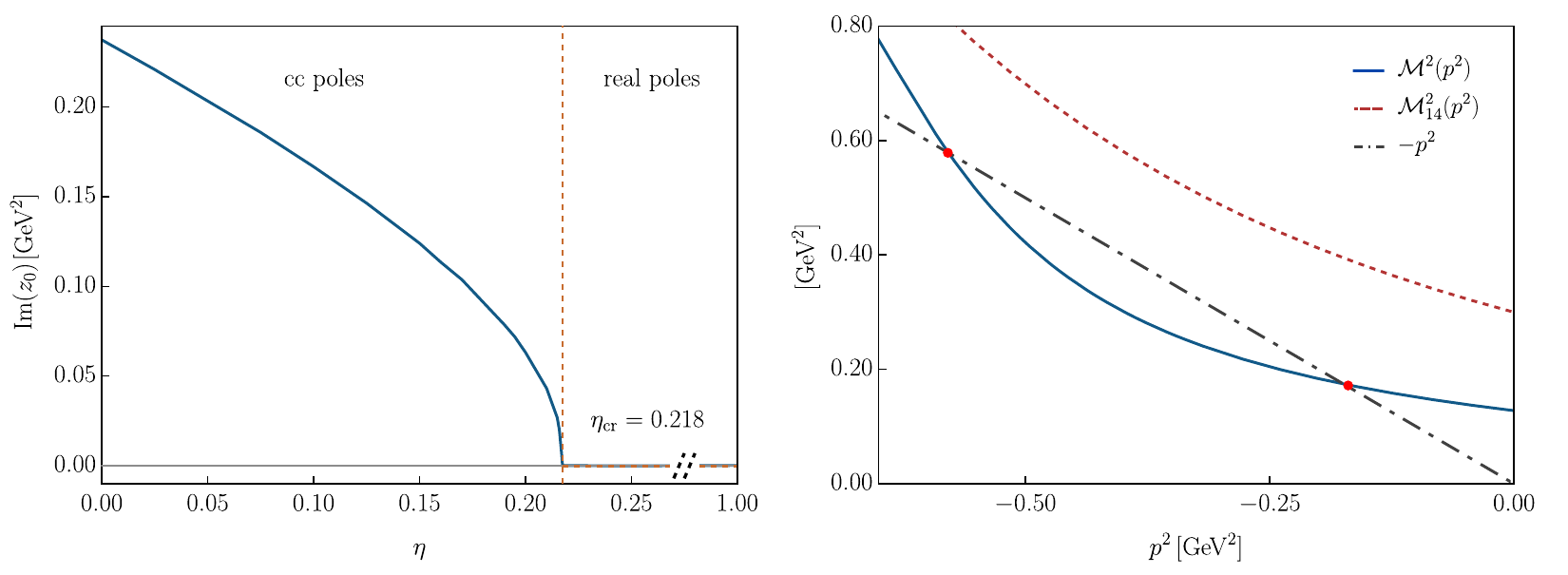}
\caption{Left panel: Imaginary part of the quark propagator pole, $z_0$, as a function of  the parameter $\eta$, \1eq{eq:geff}. Right panel: zeros of the function \mbox{$p^2+\mathcal{M}^2(p^2)$},
for two different cases, the $\mathcal{M}^2(p^2)$ 
obtained with the full quark-gluon vertex active in the gap equation (blue solid curve), and the 
$\mathcal{M}_{14}^2(p^2)$, 
corresponding to the case where only the TLC and ACM components are included (red-dashed curve).
}
\label{fig:magnet}
\end{figure*}

The position of $p_1^2$ and $p_2^2$ remains stable, varying by less than $5\%$ under changes in both the data sets and the number of points used to construct the extrapolations. Moreover, in all cases, we find no evidence of clustered non-analyticities in their vicinity, supporting their interpretation as isolated singularities.

Note that both functions $A(p^2)$ and 
${\cal M}(p^2)$ are analytic within the 
domain of the defining parabola. Thus, the 
poles observed in $\sigma_{\!\s{V}}(p^2)$ and $\sigma_{\!\s{S}}(p^2)$ originate 
from the zeros of the expression
$p^2 + {\cal M}^2(p^2)$, 
or, equivalently, the intersection 
of ${\cal M}^2(p^2)$ with the straight line $-p^2$ (see discussion at the end of \sect{sec:thetriad}).

Given the robustness of the calculated locations,~\1eq{eq:polepos}, 
and residues,~\1eq{eq:poleres}, of these two lowest-lying poles, 
their physical interpretation warrants careful analysis. 
That of the first pole at $p_1^2$ is straightforward: 
it signals a propagating mode of a massive Dirac fermion with a dynamically 
generated mass of $400$~MeV. The corresponding residue of the vector 
component $\sigma_{\!\s{V}}(p^2)$ is expected to be roughly equal to or 
slightly smaller than $1/A(0)$, following the discussion in~\cite{Wieland:2026iml}. 
According to standard field-theoretic requirements, a strict particle 
interpretation requires this value to remain bounded below unity~\cite{Itzykson:1980rh}. 
While our current numerical result is slightly larger than one, this minor 
deviation is likely an artifact of the employed truncation approximations.

In contrast, the second pole carries a strictly negative residue, 
signaling the presence of an unphysical ghost state. Consequently, the 
associated spectral functions assume negative values, establishing that the 
quark propagator violates reflection positivity. Because the underlying 
quark states do not span a positive-definite Hilbert space, quarks are 
formally prohibited from appearing as asymptotic states~\cite{Oehme:1994pv,Alkofer:2000wg}. 
Importantly, this structural feature accommodates the scenario where quark states act 
as parent states within a non-perturbative BRST quartet framework~\cite{Alkofer:2010cwc,Alkofer:2011pe}, 
providing a concrete, real-axis realization of a central mechanism of color 
confinement~\cite{Kugo:1979gm,Alkofer:2006fu}.

At this point, one might speculate that the exact quark propagator in 
any linear covariant gauge possesses an infinite tower of poles on the 
time-like half-axis. These states would feature alternating-sign residues 
with decreasing absolute values at high $-p^2$, a mathematical configuration 
necessary to satisfy the axiomatic condition that the propagator vanishes 
as $p^2 \to \infty$ in all directions of the complex plane~\cite{Oehme:1994pv}.

Independent of this conjecture, our concrete numerical results provide 
clear evidence that the unquenched quark propagator in the Landau gauge 
possesses exclusively real poles while actively violating positivity. To the 
best of our knowledge, this real-axis mechanism offers a fundamentally novel 
perspective on the functional realization of color confinement.

\subsection{The interplay between ACM and SMC}\label{sec:thetriad}

In order to gain further insights on 
the origin of these poles,
as well as the 
conditions that drive the transition 
from complex conjugate to real,  
we have analyzed the impact of 
individual form factors $\lambda_i(p^2)$
on the analytic structure of the quark propagator. 

The first observation is that 
the combined contribution of $\lambda_1(p^2)$, $\lambda_4(p^2)$, 
and $\lambda_7(p^2)$ already captures the essential features 
of $S(p)$. The remaining five form factors produce measurable changes in 
the pole position, especially of $p^2_{2}$, but do not qualitatively 
modify the basic picture. 

We therefore focus on the analytic structure of $S(p)$ generated by the 
subset of form factors $\{ \lambda_1, \lambda_4, \lambda_7\}$,
as they are sequentially activated. 
To understand the distinct impact of these three form factors, it 
is illustrative to first examine their individual contributions to the 
quark propagator. To this end, the quark gap equation is solved by retaining 
only these specific vertex components. Remarkably, yet in complete 
agreement with the recent study of~\cite{Wieland:2026iml}, the TLC~$\lambda_1$ 
provides only a minor contribution. In the upper panels of \fig{fig:dress}, the individual impacts of the TLC, ACM, and SMC on the 
propagator functions $B(p^2)$ and $A(p^2)$ are displayed. As clearly shown in 
the upper left panel, the dominant contribution to the function $B(p^2)$---and 
therefore the primary mechanism driving the dynamical generation of the quark mass---%
originates from the ACM~$\lambda_4$. 
Notably, the SMC~$\lambda_7$ does not contribute to $B(p^2)$ at all. Conversely, 
the largest contribution to the dynamical dressing of $A(p^2)$, and thus to its 
momentum dependence, is driven by the SMC. Therefore, the SMC form factor 
exerts a significant influence on the momentum dependence of 
${\cal M}(p^2) = B(p^2)/A(p^2)$ without directly participating in the 
dynamical mass generation.

Together, the ACM and SMC form factors induce a distinct momentum 
dependence in the mass function ${\cal M}(p^2)$ compared to the 
standard result originating from the TLC alone, which is the sole component 
retained in the RL truncation. This difference is most 
apparent in the product $p^2 {\cal M}(p^2)$, displayed in the lower left panel of 
\fig{fig:dress} alongside the corresponding quantity from an RL-truncated 
gap equation (black line). The dynamical part of the RL result is primarily 
concentrated below $2$~GeV; above this threshold, the mass function exhibits a 
characteristic $1/p^2$ behavior with logarithmic corrections. In contrast, the 
results incorporating the ACM and SMC display significant dynamical 
contributions extending up to several GeV. Specifically, the solid red line shows 
$p^2 {\cal M}(p^2)$ with only the TLC and ACM structures retained, while the solid 
blue line includes the full contribution of the TLC, ACM, and SMC. Common to 
both curves is a drastically modified ultraviolet behavior: the dominant 
contribution to the mass function in the several GeV range is a $1/p^4$ term, in 
excellent agreement with the parametrization provided in~\cite{Oliveira:2025boh} 
for their lattice data. Further details regarding this momentum dependence can be 
found in~\cite{Wieland:2026iml}.

Our results, which further confirm the respective findings 
of~\cite{Oliveira:2025boh,Wieland:2026iml}, make it evident that at Euclidean 
momenta around $2$~GeV, the quark mass function still contains substantial 
non-perturbative contributions. This physical feature motivates our choice of 
the renormalized mass $m_R = 50$~MeV at the renormalization scale $\mu = 2$~GeV. 
Deep within the perturbative domain, the mass function decreases to a few MeV, 
thereby accurately representing a physical up quark.

The lower right panel of \fig{fig:dress} displays a comparison of our calculated quark 
mass function with the parametrization of the lattice data provided 
in~\cite{Oliveira:2025boh}. We attribute the remaining mismatch to the use of 
the soft-gluon approximation within the quark-gluon vertex truncation. 
Nevertheless, it is evident that this momentum dependence yields a 
significantly closer agreement with the lattice data than the result 
obtained from a standard RL truncation.

Next, we analyze the impact of the individual quark-gluon vertex 
form factors on the locations of the two lowest-lying poles. 
The inclusion of the TLC~$\lambda_1$ alone in the gap equation gives rise to 
a low value for ${\cal M}(0)$ of about $180$~MeV, while maintaining an 
analytic structure characterized by two real poles. This behavior is 
expected from studies of the quark gap equation featuring subcritical or 
marginally critical couplings~\cite{Pawlowski:2024kxc,Alkofer:2026vux}. 
As the ACM~$\lambda_4$ is gradually introduced, dynamical mass generation 
increases. However, when the contribution of the ACM is turned up to 
approximately $35\%$ of its full strength, the two real poles merge; 
for even greater strengths of the ACM, they evolve into a pair of 
complex conjugate poles.

The full addition of the ACM $\lambda_4$, 
\ie taking into account the doublet $\{ \lambda_1, \lambda_4\}$,
provides sufficient strength to increase 
${\cal M}(0)$ to 546 MeV; however, now the two lowest-lying poles evolve into a complex conjugate pair, located at 
$z_0$ and $\bar{z}_0$.
Then, the inclusion of $\lambda_7$ 
brings ${\cal M}(0)$  to $310$~MeV, 
which is close to the final value of 
${\cal M}(0) = 350$ MeV, obtained when
all eight form factors are in. In addition, and most importantly, 
the poles of the quark propagator become real again, and remain so  
after the inclusion of the subleading form factors
$\lambda_i$, $i=2,3,5,6,8$.\footnote{Note that the present  described mechanism verifies a conjecture recently formulated in \cite{Pawlowski:2024kxc}.} 
This general picture is succinctly summarized 
in \tab{tab:mass}.

\begin{table}[!t]
    \centering
    \begin{tabular*}{\linewidth}{p{0.33\linewidth}|p{0.3\linewidth}|p{0.25\linewidth}}
    \toprule\toprule
    \centering quark-gluon vertex & 
    \centering poles & 
    \centering $\mathcal{M}(0)~[\textrm{MeV}]$ \tabularnewline
    \midrule
    \centering $\lambda_1$                     & \centering real & \centering $180$ \tabularnewline
    \centering $\lambda_1,\lambda_4$           & \centering cc  & \centering $546$ \tabularnewline
    \centering $\lambda_1,\lambda_4,\lambda_7$ & \centering real & \centering $310$ \tabularnewline
    \centering \raisebox{-0.01cm}{all form factors}                & \centering \raisebox{-0.01cm}{real} & \centering \raisebox{-0.01cm}{$350$} \tabularnewline
    \bottomrule\bottomrule
    \end{tabular*}
    \caption{The type of the two lowest-lying poles of the quark propagator and the value of the constituent quark mass, $\mathcal{M}(0)$, for the different sets of the quark-gluon vertex form factors retained in the gap equation (cc=complex conjugate).}
    \label{tab:mass}
\end{table}

In order to explore in detail 
the transition from 
the complex conjugate to real poles, we introduce 
an effective quark-gluon vertex, $\gt^{\,\mu}_{\!\eta}$, given by 
\be\label{eq:geff}
\gt^{\,\mu}_{\!\eta}(q,p,-k)\!=\!\lambda_1^{sg}(p^2)\bar{\tau}^\mu_{\!1} +\lambda_4^{sg}(p^2)\bar{\tau}^\mu_{\!4}+\eta\lambda_7^{sg}(p^2)\bar{\tau}^\mu_{\!7}
\,,
\ee
with $\eta \in [0,1]$. Evidently, the 
extreme case $\eta =0$ corresponds to 
the inclusion in the gap equation of the 
doublet $\{ \lambda_1, \lambda_4\}$, while 
$\eta =1$ describes the impact of the full triplet 
$\{ \lambda_1, \lambda_4, \lambda_7\}$.
The role of the parameter $\eta$ is to 
illustrate how the nature of the poles changes as one gradually increases the strength of  
$\lambda_7$.  

In \fig{fig:poles_eta} we display the pole structure of 
$\sigma_{\!\srm{V}}(p^2)$ for four representative values of the 
coefficient $\eta$. As $\eta$ is tuned towards its physical value, 
$\eta=1$, the analytic structure undergoes a qualitative transformation.

Specifically, at $\eta=0$ (upper-left), the contribution from 
$\lambda_7(p^2)$ vanishes, and the quark propagator exhibits a pair 
of complex conjugate poles, \mbox{$z_0=(-0.4+0.23 i)~\textrm{GeV}^2$}, and 
\mbox{$\bar{z}_0=(-0.4-0.23 i)~\textrm{GeV}^2$}. As $\eta$ increases, 
thereby gradually restoring the contribution from $\lambda_7(p^2)$, the 
poles approach one another (upper-right). A transition occurs at the 
critical value $\eta_{\rm cr}=0.218$, where the two poles merge into a 
single pole of order two  (lower-left). Beyond this threshold, the impact 
of $\lambda_7(p^2)$ becomes sufficiently strong to counteract $\lambda_4(p^2)$, driving the poles on the real axis into two separated simple poles (lower-right).

This behavior is further illustrated in  the left panel of \fig{fig:magnet}, where we plot the imaginary part of $z_0$, denoted by 
${\rm Im}(z_0)$, as a function of 
the tuning parameter $\eta$. 
The ${\rm Im}(z_0)$
decreases monotonically with increasing $\eta$, and vanishes exactly for 
$\eta=\eta_{\rm cr}$. Past this critical point, the pole remains purely 
real.
The resulting curve in \fig{fig:magnet} resembles the order parameter 
behavior in a second-order phase transition, such as, e.g., temperature-
dependent magnetization.

The right panel of \fig{fig:magnet} graphically displays the zeros of 
the propagator denominators, $p^2 + {\cal M}^2 (p^2)$, for time-like 
momenta up to $0.8$~GeV. Here, ${\cal M}_{14}(p^2)$ denotes the solution 
incorporating only the TLC and the ACM. This curve does not intersect the 
$-p^2$ line, yielding no real solution to the pole condition $p^2 + {\cal M}^2 (p^2)=0$. 
Conversely, when including the TLC, ACM, and SMC, the resulting mass function 
produces exactly two real zeros. Two remarks are in order regarding this mechanism: 
First, the drastically altered momentum behavior on the Euclidean side plays a 
decisive role, aligning with the analytical model for quark propagator poles 
provided in~\cite{Wieland:2026iml}. Second, and more importantly, the orientations 
of these crossings in the right panel of \fig{fig:magnet} are geometrically 
distinct. Consequently, the opposite signs for the residues emerge naturally 
from this topological analysis.

\section{Conclusions}\label{sec:conc}

In this work we have studied the 
coupled system of functional 
equations that 
describes the up quark propagator and 
the quark-gluon vertex in the complex plane. 
The structure of the vertex SDE 
employed is dictated by the 
symmetry-preserving approach  
developed in \cite{Miramontes:2025imd,Ferreira:2025wpu,Ferreira:2026gbe}, 
and furnishes the momentum evolution 
of all eight vertex form factors, in the Landau gauge.   
The analysis is simplified by evaluating these form factors in the 
soft-gluon limit, where they become a function of a single 
kinematic variable.

Our numerical results indicate that the nature of the poles in the quark 
propagator depends crucially on the structure of the quark-gluon vertex 
employed in the gap equation. When the kernel is composed solely of 
the ``classical'' component $\lambda_1$, the dynamical mass generation 
is much too weak, and the quark propagator displays two low-lying real poles, 
typical of subcritical or marginally critical coupling strengths.

With a kernel containing the ``classical''
 and the ``anomalous chromomagnetic moment'' components,
$\lambda_1$ and $\lambda_4$, the quark 
 propagator develops complex conjugate 
 poles. Instead, when the kernel 
 receives contributions also from the 
 ``spin-momentum curvature'' component,
 $\lambda_7$, the poles become real again,
 possessing residues with opposite algebraic signs.
This characteristic structure persists 
 even after all remaining 
 form factors have been included into the gap equation.

The alternating signs of the residues imply a positivity-violating 
quark propagator, a direct consequence of the negative domains induced in 
the spectral function by the ghost pole. Consequently, the space of quark 
states is not positive-definite. This formally precludes the possibility 
of quarks acting as BRST singlets, requiring them instead to be part of a 
non-perturbative BRST quartet~\cite{Alkofer:2000wg}. This structural mechanism 
constitutes a core aspect of color confinement.

 Future extensions, most notably moving beyond the soft-gluon limit in 
the evaluation of the quark-gluon vertex form factors, are planned to 
further corroborate and elucidate the insights obtained here into the 
analytic structure of the quark propagator. Furthermore, implications for 
hadron physics will be investigated to seek empirical evidence for these 
quark-level mechanisms. In this respect, it would be highly interesting to 
explore whether the framework presented here can serve as a firm foundation 
for understanding the phenomenologically successful $^3P_0$ model of 
meson decays, as suggested by the discussion in~\cite{Alkofer:2023syz}.
 
\section*{Acknowledgments}\label{sec:Acknow}
A.S.M., J.M.M. and J.P. are funded by the Spanish MICINN grants PID2020-113334GB-I00 and PID2023-151418NB-I00, the Generalitat Valenciana grant CIPROM/2022/66, and CEX2023-001292-S by MCIU/AEI. J.M.M. also acknowledges support from UV and Banco Santander, S.A. (Santander Investigación Postdoctoral program). The computations have been carried out at the CEAFMC and UHU High Performance Computer (HPC@UHU), funded by FEDER/MINECO project UNHU-15CE-2848, and on the General Computing Infrastructure (GLUON) of the UV.

\bibliographystyle{elsarticle-harv}
\bibliography{bibliography.bib}

\end{document}